\documentstyle[prl,aps,multicol,epsfig]{revtex}
\begin{document}
\title{Harmonic generation from $\rm YBa_2Cu_3O_{7-\delta}$ microwave resonators}
\draft
\author{Durga P. Choudhury${}^{1,2}$,John S. Derov${}^2$ and S. Sridhar${}^1$
\footnote{e-mail: srinivas@neu.edu, Web: http://sagar.physics.neu.edu/}}
\address{${}^1$Department of Physics, Northeastern University, Boston, MA 02115}
\address{${}^2$Air Force Research Laboratory, Hanscom Air Force Base, Bedford, MA 01730}
\date{\today}
\maketitle
\newcommand{\T}{\textstyle}
\begin{abstract}
Measurement of harmonic power on a suspended stripline microwave resonator
patterned out of thin film $\rm YBa_2Cu_3O_{7-\delta}$ (YBCO)
has been carried out as a function of temperature
and microwave power. The third harmonic power $P_3$ shows a subtle nonlinear dependence
on the fundamental power $P_1$ on a log-log scale. Fit to a straight line yields slopes
of $P_3$ {\it vs.\/} $P_1$
between 1.5 to 3, contrary to elementary calculations that predict a slope
of exactly 3. It is shown that third harmonic power generated from higher order terms
in the nonlinear impedance could account for the discrepancy quantitatively
{\em without} resorting to
any additional ad-hoc assumptions. Results of measurement of fifth harmonic power
is presented as a confirmation. A small second harmonic power is also observed that
does not show anomalous power dependence within the experimental accuracy.

\end{abstract}
\begin{multicols}{2}
Since the day of its discovery, the new breed of oxide superconductors with transition
temperatures above the boiling point of nitrogen have shown great potential as a replacement
for metals in passive microwave structures owing to their very low surface resistance
\cite{DPChoudhury98a,MJLancaster,ZYShen94b}. However, widespread applications have been hampered by the
inherent nonlinearity that these materials display. The nonlinearity
manifests itself as an increase of insertion loss of the device with power level as well as harmonic
generation and intermodulation distortion. A systematic understanding of the phenomenology
and microscopic origin of the nonlinear response is therefore very important for
successful device applications.

Measurement of the third order intercept (TOI), defined as the input power at which
the fundamental and the third harmonic output power are equal to each other, is a quantitative
measure of the nonlinear response of the device. The TOI can be measured both via a
single tone (harmonic generation) as well as multi-tone (intermodulation distortion)
measurement. Usually the former is carried out in a nonresonant transmission line
structure and the later in a high-$Q$ resonant structure with the applied tones close
to the resonance frequency $f_0$. While the single tone experiment has the simplicity of
design (only one signal source is needed), it lacks the sensitivity of the multi-tone
measurement. The multi-tone measurement has the drawback of additional complexity
owing to the need of multiple signal sources, power combiners and carefully
matched waveguides. In the experiment described in this paper, we have carried
out single tone measurements on a resonant structure, combining the advantage of
both techniques. The problem of single-tone measurement on a resonant structure, namely,
the shift of resonant frequency due to any change in experimental parameters was
solved by dynamically tuning the synthesized source before taking each individual
data point. Since $\Delta f_0=f_0(P_{\mbox{min}})-f_0(P)\propto\Delta X_S$
where $X_S=2\pi\mu_0f_0\lambda$ is the surface reactance, the effect of
microwave power induced change to the kinetic inductance of the device is also
measured simultaneously.

The experiments were carried out on films patterned out of
five different YBCO wafers on $\rm LaAlO_3$ substrate obtained commercially.
The wafers were patterned using
standard photolithography techniques (wet etching) into straightlines $100\:\mu m$
wide and $0.4\:\mu m$ thick.
Using a standard microwave formula, the characteristic
imedance $Z_0$ was evaluated to be $72\:\Omega$. The patterned structure
has YBCO only on one side, i.e. it does not use superconducting ground planes. The
copper package housing the film also provides the necessary grounding. This
gives us the necessary sensitivity to measure the effect of very small applied DC
magnetic fields which would be shielded out by a superconducting ground plane.
Such a capability is important and unique to our setup as small DC fields have
been shown to {\em improve} device characteristics\cite{DPChoudhury96a}.
The results of the field measurements, kinetic inductance effects, details of
temperature dependence and some
related topics will be described elsewhere\cite{DPChoudhury99b}.

The resonators ($f_0=4.3\:$GHz) were characterized using a vector network analyzer.
The unloaded $Q$ was $\sim 1.2$-$1.8\times 10^4$ at 10\,K. Coupling was subsequently
increased to transmit maximum power through the resonator. The nonlinearity manifests
itself as a distortion of the Lorentzian shape of the resonance curve\cite{BAWillemsen95a}
and a decrease of $Q$.

The synthesizer was set to generate a CW signal at the resonance frequency that
was amplified by 30\,dB, filtered by a coaxial 5\,GHz low pass filter and was fed
to the resonator. The output of the resonator was analyzed with a spectrum analyzer.

The aspect of harmonic generation or intermodulation distortion that is most commonly
investigated is its dependence on the input power. The knowledge of this dependence
is very important both from the point of view of device design as well as understanding
the mechanism of nonlinearity. A simple algebraic argument, as shown below, predicts
the third order nonlinear power to scale as the cube of the input power. However, there
is a lack of consensus about this in the experiments carried out by different investigators.
Not only does the dependence differs from a cube\cite{CWilker95a,GHampel97a}, it also depends upon
the input power. It is often seen that the dependence is closer to $P_{3\omega}\propto P_i^2$
rather than $P_i^3$\cite{BAWillemsen99b}.

In this letter, we show that this discrepancy comes about because of:
\begin{itemize}
\item Failure to take into account the power dependence of the insertion loss of
the device.
\item The assumption that the entire third harmonic power comes from the cubic term
$\partial^2Z_S/\partial i^2$ of the current-dependent nonlinear impedance $Z_S$.
\end{itemize}

The first point, often neglected in the analysis of nonlinear frequency conversion
data is that the harmonic power is generated by the current flowing in the
device which does not scale quadratically with the {\em input} power because of
the power dependence of the insertion loss.
In the case of a moderately nonlinear device this manifests as a distortion of the
Lorentzian shape of the resonance in the frequency domain
corresponding to the power being transfered to higher harmonics. In the
extreme case, this can lead to saturation and discontinuous
changes in output power. Higher levels of applied power go to heating the
device {\em decreasing} the output power. Fig.\,\ref{Trace} shows the value of
the transmission amplitude $S_{12}$ for one of the resonators
obtained from a vector network analyzer as the frequency is swept about the resonance
at input power levels of 10, 5, 0, $-5$ and $-15$\,dBm respectively, 10\,dBm corresponding
to the lowest trace and $-15$\,dBm corresponding to the highest. Decreasing the input
power below $-15$\,dBm value did not increase $S_{12}$ any further, indicating that the
device is operating in the fully linear regime. Similarly, increasing the input power
above 10\,dBm did not seem to lower the saturation any futher. Depending on the sample,
other manifestations of nonlinear behavior, such as a step-like change superimposed on
a Lorentzian $S_{12}$ trace were also observed. Many of these can be modeled as a lumped
resonator involving nonlinear elements\cite{JHOates93a}. Extremes of such nonlinear
behavior also serve as a characterization of the quality
of the superconducting film as more granular films tend to demonstrate this behavior.
{
\narrowtext
\begin{center}
\begin{figure}
\mbox{\epsfig{file=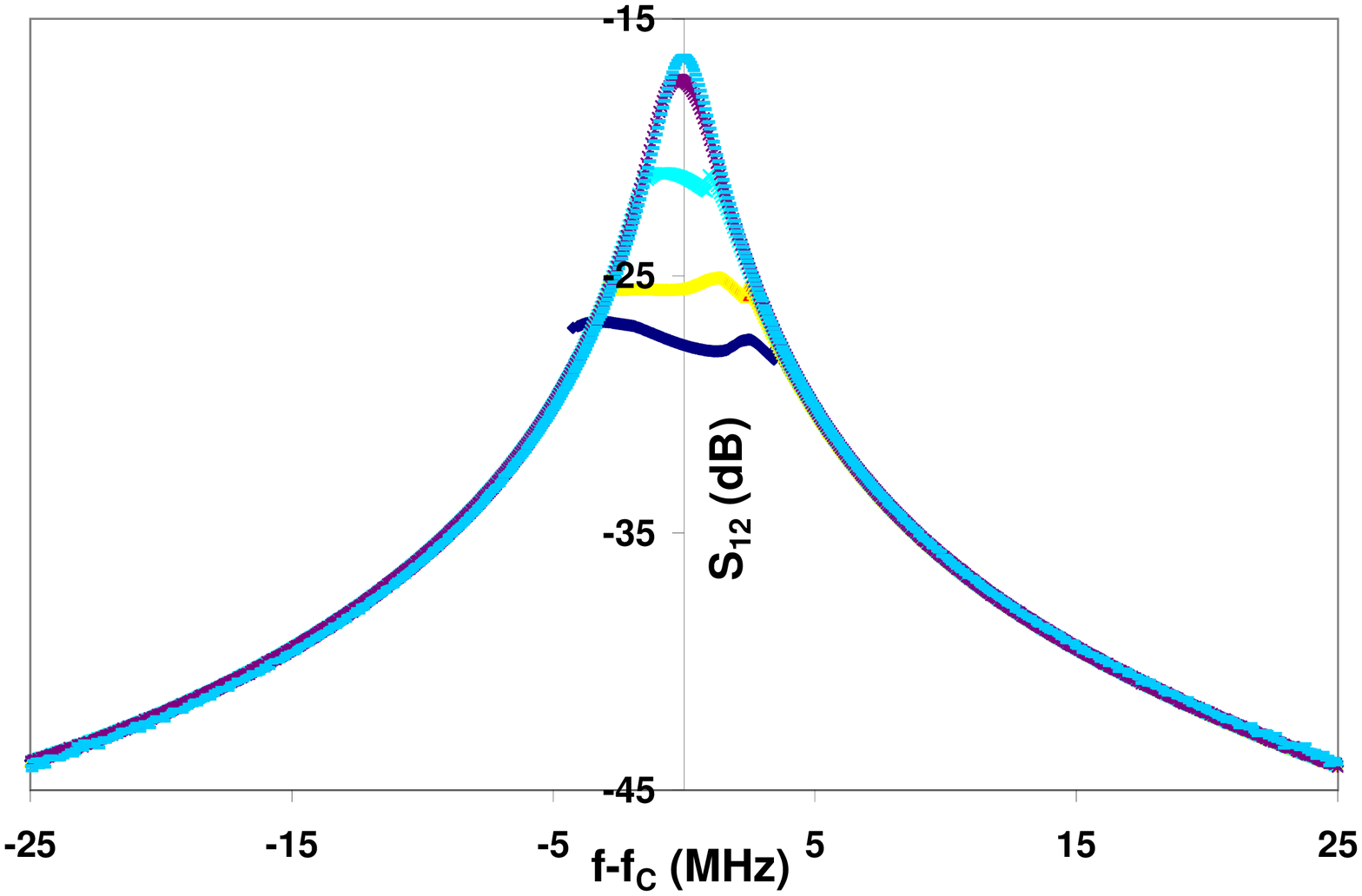, height=2.4 truein, width=0.45\textwidth}}
\caption{Transmitted power {\it vs.\/} frequency about the resonance frequency at
input powers of 10, 5, 0, -5 and -15 dBm, from bottom to top.}
\label{Trace}
\end{figure}
\end{center}
}
Before each measurement of harmonic power reported below,
the resonance was characterized using a network analyzer to ensure that the input/output
antennas were not overcoupled and to verify that the response of the resonator was still
the same. Direct coupling
between the antennas were found to be at least
50\,dB below the resonance, ensuring that the transmitted power comes from
microwave currents induced in the resonator only.

It is therefore clear that trying to calculate the slope of the harmonic power against
the input power can give misleading results. One way to correct for the error is
to calculate the RMS current in the resonator taking into account the insertion loss
of the device\cite{DEOates90a}. In this paper we plot the harmonic power against
the {\em output} power at the fundamental frequency which automatically compensates
for insertion losses.

The second implicit assumption made in the analysis of the dependence of the third
harmonic power on the input power is that the entire contribution to the former
comes from
the cubic term in the nonlinear impedance. However, any higher order odd term will
also contribute to the third harmonic. Assuming $v=ir_1+i^3r_3+i^5r_5$, $i=i_0\sin\omega t$
the third harmonic term can be written as $v_{3\omega}
={\T -r_3i_0^3\over\T 4}\left(1+{\T 5\over\T 4}i_0^2{\T r_5\over\T r_3}\right)$, the
negative sign implying that
the 3rd harmonic is phase shifted by $\pi$ from the fundamental. Since spectrum analyzer
measurements are not sensitive to relative phase, we ignore the negative sign.
Choosing a reference value for current $i_r$ (for example, the current that
generates 1\,mW of power in case of a dBm scale) and
taking logarithm of both sides and assuming ${\T r_5i_0^2\over\T r_3} \ll 1$, we get
\begin{equation}
\log {\T 4v_{3\omega}\over\T r_3i_r^3}=3\log {\T i_0\over\T i_r}+{\T 5\over\T 4}i_0^2{\T r_5\over\T r_3}
\label{eq1}
\end{equation}
The presence of a fifth (or higher) order contribution to the third harmonic power would
be indicated by a simultaneous observation of the fifth harmonic signal.

Typical data for second, third and fifth harmonic of one of the films at 30\,K is shown
in fig.\,\ref{Harm}.
The inset shows second and third harmonic data of a different film from a different
manufacturer at 67.5\,K. The second harmonic power is much weaker than the third (and
even the fifth) as expected of a superconductor. The fifth harmonic power
falls below the noise floor of the spectrum analyzer at a fundamental power of $\sim -8$
dBm because the noise floor of the spectrum analyzer is significantly higher at higher
frequencies.
The second harmonic power, being very weak, is difficult to measure and analyze reliably.
In the cases where a reasonably strong second harmonic signal was seen, its slope against
the fundamental was close to 2:1 on a log-log scale, as is expected.
However, the third
harmonic power has a much more complex dependence on the input power.
Fitting the plot to a straightline yields a slope of anywhere between 1.5 to 3 
depending on the film and the temperature with a variance
of a few \%. Usually such values of the variance signifies that the fit is very good.
However, as fig\,\ref{Harm} shows, the data has a richer structure than a straighline.
{
\narrowtext
\begin{center}
\begin{figure}
\mbox{\epsfig{file=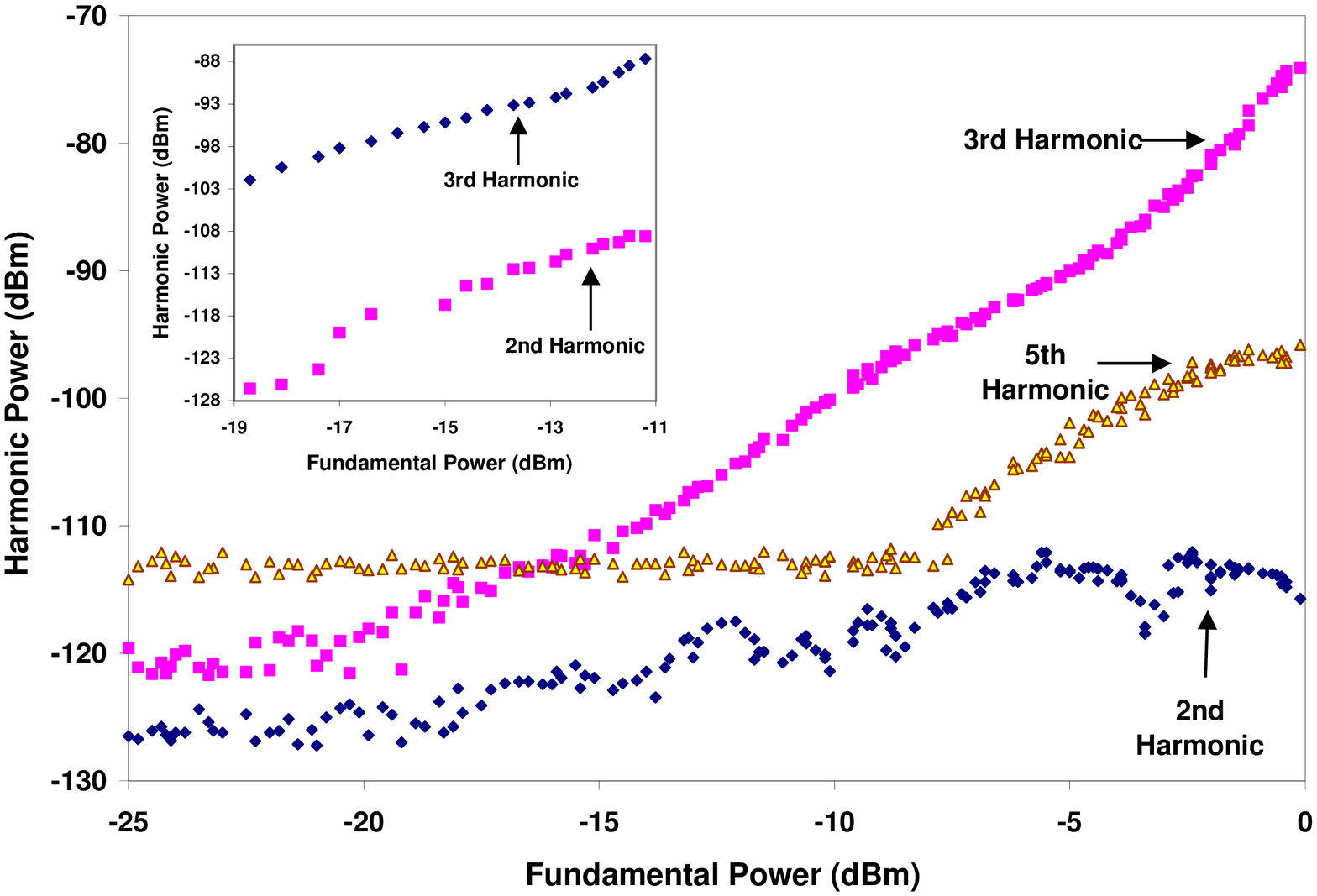, height=2.4 truein, width=0.45\textwidth}}
\caption{Typical second, third and fifth harmonic power data at 30\,K.
Inset: Second and third harmonic power of a different film at 67.5\,K}
\label{Harm}
\end{figure}
\end{center}
}
It is therefore obvious that the third harmonic power cannot be coming from the third order nonlinear
term only, as has traditionally been assumed. We can fit the data quantitatively by keeping one
to two more term in the Taylor series expansion of the nonlinear impedance, depending on the
sample. Fig.\,\ref{Fit}
shows the data of fig.\,\ref{Harm} along with a straightline fit (with a slope of 2.3118 and
variance of 0.0115) and a fit calculated by
keeping the third harmonic contribution from a fifth and a seventh order term. The second
fit is nearly indistinguishable from data. The top inset shows third harmonic data of a different
film at 55\,K, a straightline fit (with slope of 2.48 and variance of 0.1142) and a fit keeping
upto seventh order terms. The lower inset shows the second harmonic data of the same film at
the same temperature as the upper inset and a straightline fit with a slope of 2.001 and variance
of 0.04). In those films (such as the one in fig.\,\ref{Harm})
where a seventh order term was necessary to obtain a satisfactory fit, one should be able to
see a seventh harmonic signal as well. We were, however, unable to measure it as it was outside
the bandwidth of our spectrum analyzer. A similar method of analysis has also been used by other
investigators\cite{GHampel97a}. However, to our knowledge, this is the first measurement of
the fifth harmonic which validates the argument.
{
\narrowtext
\begin{center}
\begin{figure}
\mbox{\epsfig{file=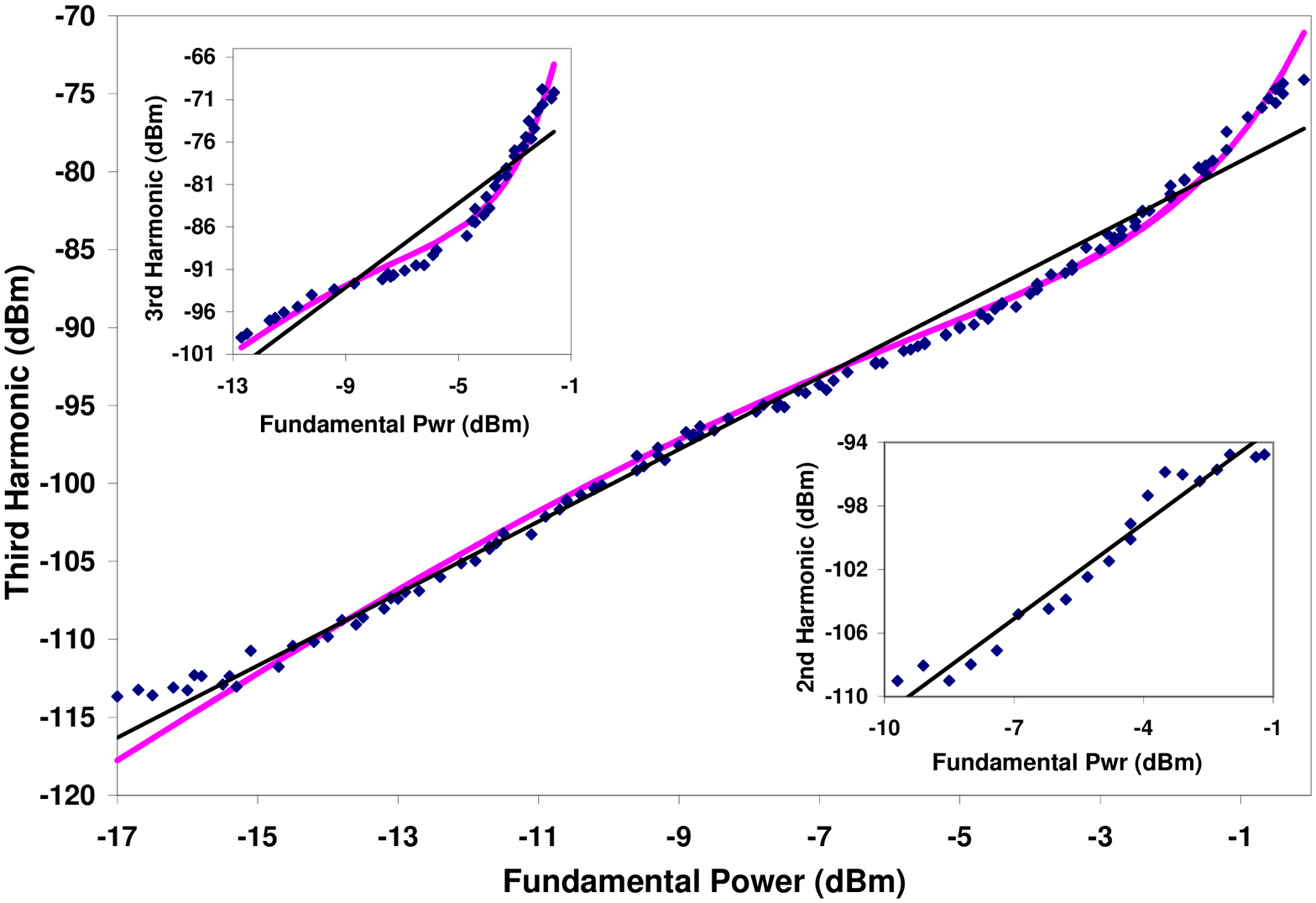, height=2.4 truein, width=0.45\textwidth}}
\caption{The same data as in fig.\ref{Harm}. A straightline fit and one
keeping a 5th and 7th order contribution is also shown. Inset: (Top) Third
and (Bottom) Second harmonic data for a different film at 55\,K along with fits.}
\label{Fit}
\end{figure}
\end{center}
}
In principle the above type of analysis should be applicable to the second harmonic
power also. However, since the measured second harmonic data fits reasonably well
to the predicted 2:1 behavior, it is likely that the contribution to the second
harmonic power from higher harmonics is small. In any case, our low pass filter
had a harmonic passband that coincided with the fourth harmonic frequency. We were,
therefore, unable to measure this quantity. Let us point out here that the even harmonics
do not contribute to the third harmonic power and odd harmonics do not contribute to
the second harmonics. Physically speaking, this implies that time reversal symmetry
breaking and preserving mechanisms of harmonic generation are independent of each
other.

There are a few mechanisms for the observed second harmonic power, {\it e.g.\/}
Josephson vortices, intrinsic mechanisms breaking time reversal
symmetry or magnetic effects superimposed on the superconducting state.
Harmonic generation experiments, being macroscopic in nature, cannot
distinguish between these mechanisms and it is quite likely than several different
mechanisms contribute to even harmonics simultaneously.
\section{Acknowledgements}
Stimulating discussions with D. E. Oates are thankfully acknowledged. Work at Northeastern
was supported by NSF-9711910 and AFOSR.
\end{multicols}
\bibliographystyle{prsty}
\bibliography{/shares/u-drive/durga/papers/strings,/shares/u-drive/durga/papers/big}
\end{document}